\documentclass[aps,prb,superscriptaddress,twocolumn,no-footinbib,showpacs,floatfix,amsmath,amssymb]{revtex4-1}
 
\usepackage{graphicx}% Include figure files
\usepackage{dcolumn}% Align table columns on decimal point
\usepackage{bm}% bold math
\usepackage[normalem]{ulem}
\usepackage{subfigure} %Subfigure
\usepackage{soul,xcolor}
\usepackage{siunitx}
\usepackage{braket}
\usepackage{textcomp}
\setstcolor{red}
\RequirePackage{color}

\usepackage[capitalise,nameinlink,noabbrev]{cleveref}
\crefname{equation}{}{}
\crefname{enumi}{}{}

\begin{document}

\title{Kekulé Induced Valley Birefringence and Skew Scattering in Graphene}% Force line breaks with \\
\author{Elias Andrade}
\email{eandrade@estudiantes.fisica.unam.mx}
\affiliation{
 Posgrado en Ciencias Físicas, Instituto de F\'{i}sica, Universidad Nacional Aut\'{o}noma de M\'{e}xico (UNAM). Apdo. Postal 20-364, 01000 M\'{e}xico D.F., M\'{e}xico
 }%
%\author{Alex Santacruz}
%\email{elias.andrade@uabc.edu.mx}
%\affiliation{%
% Facultad de Ciencias, Universidad Aut\'{o}noma de Baja California, Apdo. Postal 1880, 22800 Ensenada, Baja California, M\'{e}xico. }%
\author{Ramon Carrillo-Bastos}
%\email{ramoncarrillo@uabc.edu.mx}
\affiliation{%
 Facultad de Ciencias, Universidad Aut\'{o}noma de Baja California,
Apdo. Postal 1880, 22800 Ensenada, Baja California, M\'{e}xico. 
}%
\author{Mahmoud M. Asmar}
%\email{masmar@kennesaw.edu}
\affiliation{%
Department of Physics, Kennesaw State University, Marietta, Georgia 30060, U.S.A.
}%
\author{Gerardo G. Naumis}
%\email{naumis@fisica.unam.mx}
% \homepage{\\http://www.fisica.unam.mx/personales/naumis/}
\affiliation{
 Depto. de Sistemas Complejos, Instituto de F\'{i}sica, Universidad Nacional Aut\'{o}noma de M\'{e}xico (UNAM). Apdo. Postal 20-364, 01000 M\'{e}xico D.F., M\'{e}xico
 }%

\date{\today}% It is always \today, today,

\begin{abstract}
In graphene, a Kekul\'e-Y bond texture modifies the electronic band structure generating two concentric Dirac cones with different Fermi velocities lying in the $\Gamma$-point in reciprocal space. The energy dispersion results in different group velocities for each isospin component at a given energy. This energy spectrum combined with the negative refraction index in p-n junctions, allows the emergence of an electronic analog of optical birefringence in graphene. We characterize the valley birefringence produced by a circularly symmetric Kekulé patterned and gated region using the scattering approach. We found caustics with two cusps separated in space by a distance dependent on the Kekulé interaction and that provides a measure of its strength. Then, at low carrier concentration we find a non-vanishing skew cross section, showing the asymmetry in the scattering of electrons around the axis of the incoming flux. This effect is associated  with the appearance of the valley Hall effect as electrons with opposite valley polarization are deflected towards opposite directions.
\end{abstract}

\maketitle
\section{Introduction}\label{Sec_I}
The similarities between the Helmholtz and Schr\"odinger equations result in photons and electrons displaying similar wave phenomena\cite{book-analogies}. Furthermore, the propagation of electrons through the two-dimensional honeycomb arrangement of carbon atoms, known as graphene, leads to the dressing of electronic states as massless Dirac-like electronic excitations residing at opposite corners of the Brillouin zone~\cite{Klein-Katsnelson2006}, thus augmenting the analogies between the electronic and optical phenomena.
The ability to control the charge carrier’s group velocity via graphene gating~\cite{FieldEffect-graphene} has led to the prediction and experimental realization of electronic Veselago lensing~\cite{Veselago-GrapheneTheory,Veselago-GrapheneExp}. The sensitivity of this lensing to the conduction electrons properties aids the detection of anisotropies and tilting of the Dirac cones~\cite{Veselago-anisotropy,Veselago-tilting}, the presence of strain~\cite{Veselago-strain}, and disorder~\cite{Veselago-disorder}. Veselago lensing also facilitates the waveguiding of electrons in p-n junctions~\cite{Waveguides-NatNano} and the emergence of caustics and cusps in circular geometries~\cite{cserti2007caustics}. Moreover, similar to optical birefringence in anisotropic crystals where the group velocity depends on light polarization~\cite{BERRY1980257}, spin birefringence for electrons emerges in graphene due to the Rashba spin-orbit interaction~\cite{Rashba,Kane} which leads to distinct Fermi velocities for each spin component, and in circular geometries, spin birefringence brings about the formation of caustics with two cusps, with a space separation that depends on the strength of the Rashba spin-orbit coupling~\cite{Asmar2013-Circular,asmar2015symmetry}. 

In addition to spin, electrons in graphene possess the valley degree of freedom~\cite{graphene}. The valleys in graphene have a large separation in momentum space~\cite{katsnelson2012graphene}, which suggests that this degree of freedom can be potentially used in applications where it will play a role similar to spin in spintronics~\cite{Spintronics1,Spintronics2}. The field that aims to manipulate and control the valley degree of freedom in applications is known as {\it valleytronics}~\cite{vallfilter,valleyline,valleyline2,valleyline3,valleyline4,valleymagnetic,valleymassbarrier,valleymassbarrier2,valleyphonon,valleytrigonal,valleystrain,valleystrain1,Minimal}. Similar to spin-orbit interactions in spintronics, interactions contrasting the degenerate valleys in graphene play an essential role in valleytronincs. Such interactions include the Kekulé patterning of graphene~\cite{Mudry2007, Chamon2000},{\it i.e.}, the periodic bond modulation of the graphene lattice. Depending on the bond modulation pattern~\cite{Gamayun} two different Kekulé distortions phases can emerge: the Kekulé-Y~\cite{Gutierrez} found in graphene deposited on Cu[111] and the Kekulé-O~\cite{Bao2021,Bao2022,EOM2020,qu2022ubiquitous,StabilizingZhanng2022} that arises in bilayer graphene intercalated with Li. The tight-binding calculations by Gamayun~\textit{et al.}\cite{Gamayun} found that Kekulé-Y produces an effective interaction that leads to valley-momentum locking, while Kekulé-O leads to the formation of a gap in the electronic spectrum.

Kekulé-Y patterned graphene, breaks a valley degeneracy through valley-momentum locking which produces a low energy spectrum with two nested Dirac cones with different Fermi velocities~\cite{Gamayun}. The energy-momentum dispersion modification caused by Kekulé-Y patterning leads to drastic modifications in the magnetic and optical response of graphene~\cite{Mohammadi2022,Elias2020,Naumis2020,Herrera2020,Mohammadi_2021,Santacruz2022}, and crucially aids the control of the valley degree of freedom in the electronic transport\cite{Wang2018,Elias_2019,Tijerina_2019,Wu2020,Wang2020,Zeng2021,Galvan2022,KekSecondNeighbors}. In this paper we study the scattering of Dirac fermions from circularly Kekulé-Y-patterned regions in the semiclassical limit and we explore the effects of this interaction on electron optics and the appearance of valley birefringence. We also investigate the scattering of charge carriers in graphene from short-range scattering regions with locally enhance Kekulé-Y interactions due to adatom deposition. Our analysis of the total, transport, and skew cross sections for these short-range scatterers reveals the dependence of these cross sections on the strength of the Kekulé-Y interaction and we show the appearance of a valley Hall effects due to skew scattering from these scatterers. 

The layout of this work is as follows. In section~\ref{Sec_II} we present the model, section \ref{Sec:scattering} is devoted the scattering calculations. Valley birrefringence is analyzed in section \ref{Sec:Valley}, while in section \ref{Sec:Low} we study  the low-energy scattering. Finally, we conclude by discussing our main findings.

\section{Model}\label{Sec_II}
\begin{figure}[!htbp]
\begin{center}
\includegraphics[scale=0.25]{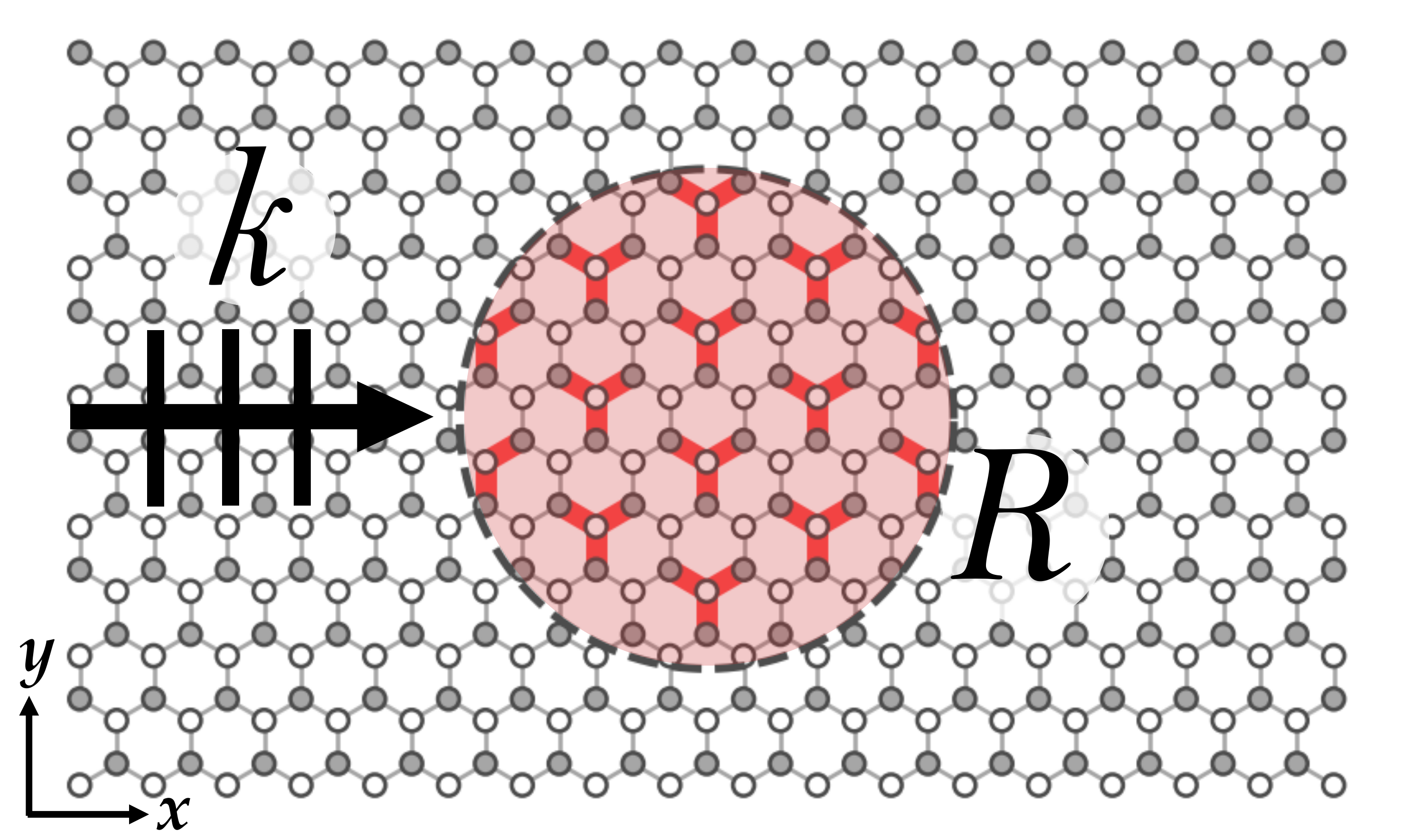}
\caption{Schematic of the system (not to scale). A graphene lattice where an incoming flux of electrons in the $x$ direction approaches a circular region of radius $R$ with a gate potential and Kekulé-Y bond texture (red bonds).}
\label{fig:system}
\end{center}
\end{figure}

Our system consist of an infinite sheet of pristine graphene containing a circularly Kekulé-ordered patch of radius $R$, Fig.~\ref{fig:system}. We consider the scattering of an incoming flux of electrons in the $x$-direction with momentum $k$. To describe the electronic properties of the  graphene sheet we adopt the low-energy description, i.e., the Dirac Hamiltonian \cite{katsnelson2012graphene}. Nevertheless, the Kekul\'e modulated portion of the lattice has a larger unit cell than non-modulated graphene lattice. Hence, to match the pristine and Kekul\'e patterned graphene wave functions it is practical to use an enlarged unitary cell for the case of undistorted graphene. This is equivalent to consider the group $C^{''}_{6v}$, with a cell with six atom graphene’s unit cell, which avoids the treatment of degenerate states at two inequivalent Dirac points~\cite{Ochoa2012}. This is more clearly seen if we start with the Hamiltonian for the Kekul\'e region and then  pristine graphene appears as a limiting case.

The space dependent Hamiltonian describing the system in Fig.~\ref{fig:system} is given by
\begin{subequations}
\begin{equation}
    \mathcal{H}=\mathcal{H}_0+\mathcal{H}_Y(r) + V(r)\;,
    \label{Eq:Ham}
\end{equation}
where,
\begin{equation}
    \mathcal{H}_{0}= v_f (\bm p \cdot \bm \sigma) \otimes \tau_0
\end{equation}
is the low-energy graphene Hamiltonian with ${\bm p}=-i\hbar(\partial_x,\partial_y)$ the momentum operator,  $v_f\sim10^6$m/s the Fermi velocity, and $\sigma$, $\tau$ the sets of Pauli matrices acting on the sublattice and valley pseudo-spin spaces respectively.
\begin{equation}
    \mathcal{H}_{Y}=\Delta v_f  \sigma_0 \otimes (\bm p \cdot \bm \tau) \Theta(R-r)
\end{equation}
is the Kekulé-Y bond perturbation~\cite{Gamayun} with amplitude $\Delta$ within the circular region,
\begin{equation}
    V(r)=V_0 \Theta(R-r)\sigma_0 \otimes \tau_0,
\end{equation}
is a constant gate potential with amplitude $V_0$ in the Kekulé circular patch, and $\Theta$ is the Heaviside function. %\EA{In general when there is a discontinuity in the Fermi velocity it is necessary to symmetrize the Hamiltonian to retain the hermiticity, in the case of the Heaviside function this results in the appearance of a Dirac Delta potential at the boundary, which produces a discontinuity of the envelope wavefunction \cite{}, however this effects should not change qualitatively our results for small values of $\Delta$.}

The Hamiltonian in Eq.~\eqref{Eq:Ham} acts on the states expressed in the valley isotropic representation:
\begin{equation}
    \Psi=
        \begin{bmatrix}
    \psi_{K'}\\
    \psi_{K}
    \end{bmatrix}
    =
    \begin{bmatrix}
    -\psi_{B,K'}\\
    \psi_{A,K'}\\
    \psi_{A,K}\\
    \psi_{B,K}
    \end{bmatrix},
\end{equation}
\end{subequations}
Notice that the subindex $A$ and $B$ in $\Psi$ corresponds to each graphene's bipartite lattice while $\boldsymbol{K}$ and $\boldsymbol{K}'$ label the valley. 
For regions outside the Kekulé modulated region, the limit of pristine graphene is recovered, $\Delta=0$, thus having a $4 \times 4$ operator which represents the Dirac Hamiltonian in the enlarged unitary cell. 

\section{Scattering}
\label{Sec:scattering}
In this section we study the scattering of Dirac fermions from a circularly symmetric Kekulé patterned region. We adopt the partial waves scattering method to find the S-matrix, which requires finding and matching the eigenstates in the different scatttering regions of our system. 
For any effective theory that uses an envelope wavefunction, as is the case of the Dirac equation for graphene, the matching requires a supplemental boundary condition of the form $\Psi=M\Psi$ in order to retain the hermiticity and preserve currents. Here $M$ is a matrix containing the microscopic details and the symmetries of the problem \cite{beenakker2008colloquium,mccann2004symmetry,AsmSuperCond,AsmConductance,BasicTheory,Ilya}. Since we consider the Kek-Y bond modulation as a perturbation within the same graphene sheet, no major misalignment is expected and thus for small $\Delta$ we can consider $M$ as unitary throughout this work. Moreover, as our system possesses circular symmetry, it is natural to evaluate its eigenfunctions in polar coordinates. 
The $z$-component of  orbital angular momentum $L_z=-i \hbar \partial_\theta$ does not commute with the Hamiltonian, $[H,L_z]= i \hbar v_f (\bm \sigma \times \bm p)_z \otimes \tau_0+i \hbar v_f \sigma_0 \otimes(\bm \tau \times \bm p)_z$. On the other hand, the sum of $L_z$ and the intrinsic angular momenta associated with the valley and sublattice degrees of freedom, {\it``valley-lattice-angular momentum'' $J_z$}, is conserved and given by
\begin{equation}
        J_z=L_z+\frac{\hbar}{2}(\sigma_z \otimes \tau_0+\sigma_0 \otimes \tau_z).
\end{equation}
Here, it is important to notice that
$[H,\frac{\hbar}{2} \sigma_z \otimes \tau_0]=-i \hbar  v_f (\bm \sigma \times \bm p)_z\otimes\tau_0$, and $[H,\frac{\hbar}{2} \sigma_0 \otimes \tau_z]=-i \hbar v_f \sigma_0 \otimes(\bm \tau \times \bm p)_z$, which leads to $[H,\frac{\hbar}{2} \sigma_z \otimes \tau_0+\frac{\hbar}{2} \sigma_0 \otimes \tau_z+L_z]=0$.
%an analogous relation holds for $\bm \tau$.
We can express the eigenfunctions in their total pseudo-angular momentum basis, such that $J_z \Psi_n=n \hbar \Psi_n$, thus
\begin{equation}
    \Psi_n (r,\theta)=e^{in\theta}
    \begin{bmatrix}
    -e^{-i\theta} \Phi_{B,K'}(r)\\
    \Phi_{A,K'}(r)\\
    \Phi_{A,K}(r)\\
    e^{i\theta} \Phi_{B,K}(r)
    \end{bmatrix}.
    \label{Eq:Espinor}
\end{equation}
where $\theta=\tan^{-1}{y/x}$, and find the radial part of the wave functions by applying the Hamiltonian in Eq.~\eqref{Eq:Ham} to our spinor in Eq.~\eqref{Eq:Espinor} to get the following set of coupled differential equations,
%for the radial part within the circle we just apply the Hamiltonian in Eq. (\ref{Eq:Ham}) to our spinor in Eq. (\ref{Eq:Espinor}) an get the following set of coupled differential equations,
\begin{subequations}\label{diffrel}
\begin{equation}
    L_n^-[\Phi_{A,K'}(r)+\Delta \Phi_{A,K}(r)]=-i(\epsilon-\nu)\Phi_{B,K'}(r),
\end{equation}
\begin{equation}
    L_{n-1}^+\Phi_{B,K'}(r)-\Delta L_{n+1}^- \Phi_{B,K}(r)=-i(\epsilon-\nu)\Phi_{A,K'}(r),
\end{equation}
\begin{equation}
    L_{n+1}^-\Phi_{B,K}(r)-\Delta L_{n-1}^+ \Phi_{B,K'}(r)=i(\epsilon-\nu)\Phi_{A,K}(r),
\end{equation}
\begin{equation}
    L_n^+[\Phi_{A,K}(r)+\Delta \Phi_{A,K'}(r)]=i(\epsilon-\nu)\Phi_{B,K}(r),
\end{equation}
where,
\begin{equation}
    L_n^{\pm}=\left(\partial_r\mp\frac{n}{r}\right),
\end{equation}
\end{subequations}
here $\epsilon=E/(\hbar v_f)$, $\nu=V_0/(\hbar v_f)$. Since $L_n^{\pm}$ acts as a ladder operator for the cylindrical Bessel functions $J_n$,
\begin{equation}
    L_n^{\pm} J_n(kr)=\mp k J_{n\pm1}(kr),
\end{equation}
thus, a natural ansatz is
\begin{align}\label{anzats}
    \Phi_{A,K}(r)&=i(\epsilon-\nu)C_AJ_n(kr),\\
    \Phi_{A,K'}(r)&=i(\epsilon-\nu)C_BJ_n(kr),
\end{align}
where $C_A$ and $C_B$ are constants, and $k$ is the electron wave number. Inserting the anzats in Eq.~\eqref{anzats} in the relations in Eq.~\eqref{diffrel}, results into the exact form of the spinor solutions and determines the wave numbers  
%The radial components for sublattice $\Phi_{B,K}(r)$ and $\Phi_{B,K'}(r)$ can then be easily obtained and the the resulting dispersion is,
\begin{equation}\label{kpkm}
    k_{\pm}=\frac{|E-V_0|}{\hbar v_f(1\pm \Delta)}.
\end{equation}
Thus the n$^{\rm th}$ angular momentum eigenstates in the inner region are,
\begin{equation}
\begin{split}
    \Psi_n^<(r,\theta)=&T^+_n e^{in\theta}
    \begin{bmatrix}
     J_{n-1}(k_+ r)e^{-i\theta}\\
    is' J_n(k_+ r)\\
    is' J_n(k_+ r)\\
    - J_{n+1}(k_+ r) e^{i\theta}
    \end{bmatrix}\\
    &+T^-_n e^{in\theta}
    \begin{bmatrix}
     J_{n-1}(k_- r) e^{-i\theta}\\
    is' J_n(k_-r)\\
    -is' J_n(k_- r)\\
     J_{n+1}(k_- r) e^{i \theta}
    \end{bmatrix},
\end{split}
\end{equation} 
where $T_n^+$ and $T_n^-$ are determined by $s'=\text{sgn}(E-V_0)$ and the boundary conditions. Since the pseudo-angular momentum is conserved during the scattering process, we can treat each component of $n$ independently and use the partial wave method to determine the S-matrix elements. In the region $r>R$, we describe the wavefunction in terms of incoming (in) and outgoing (out) cylindrical waves, where the corresponding spinor for each valley is
\begin{subequations}
\begin{equation}
\psi^{(\text{out})/(\text{in})}_{n,K'}(r,\theta)\ket{K'}=
    \begin{bmatrix}
            H_{n-1}^{(1)/(2)}(kr)e^{i(n-1)\theta}\\
            isH_{n}^{(1)/(2)}(kr)e^{in\theta}
        \end{bmatrix}
        \ket{K'},
\end{equation}
\begin{equation}
\psi^{(\text{out})/(\text{in})}_{n,K}(r,\theta)\ket{K}=
    \begin{bmatrix}
            -isH_{n}^{(1)/(2)}(kr)e^{in\theta}\\
            H_{n+1}^{(1)/(2)}(kr)e^{i(n+1)\theta}
        \end{bmatrix}
        \ket{K},
\end{equation}
\end{subequations}
here $H_n^{(1)}$ and $H_n^{(2)}$ are Hankel functions of the first and second kind respectively, and $s=\text{sgn}(E)$. Now we can write the wavefunctions in terms of the scattering matrix $S_n$ such that $\psi_n=\psi_n^{(\text{in})}+S_n \psi_n^{(\text{out})}$,
\begin{equation}
\begin{split}
    \Psi_n^>(r,\theta)=
    &\sum_{\alpha} c_\alpha \psi_{n,\alpha}^{(\text{in})}(r,\theta)\ket{\alpha}\\
    & +\sum_{\alpha,\beta} c_\alpha S_{n,\alpha\beta} \psi_{n,\beta}^{(\text{out})}(r,\theta)\ket{\beta},
    \label{Eq:Psi_out_n}
\end{split}
\end{equation}
where $\alpha$, $\beta$ $\in$ \{$K$,$K'$\} and $S_{n,\alpha\beta}$ corresponds to the scattering from $\alpha$ to $\beta$ valley, $c_{K}$ and $c_{K'}$ are the weights of the valley polarization. We can obtain the coefficients for $S_n$, $T_n^+$ and $T_n^-$ by applying the boundary conditions at $\Psi_n^<(R,\theta)=\Psi_n^>(R,\theta)$, as shown in Appendix A. Additionally, an incident plane-wave in the $x$-direction can be expressed with the aid of the Jacobi-Anger expansion as,
\begin{equation}
    e^{ikr\cos\theta}=\sum_{n=-\infty}^{\infty}i^n J_n(kr)e^{in\theta},
\end{equation}
or equivalently as,
\begin{equation}
    \Phi_0(r,\theta)=\sum_{n=-\infty}^{\infty}\sum_\alpha c_{\alpha} \frac{i^n}{2}[\psi_{n,\alpha}^{(\text{out})}(r,\theta)+\psi_{n,\alpha}^{(\text{in})}(r,\theta)] \ket{\alpha}.
\end{equation}
The latter allows one to express $\Psi^>(r,\theta)$ in terms of the incoming plane and the outgoing  waves, {\it i.e.}
\begin{equation}
\begin{split}
    &\Psi^>(r,\theta)= \Phi_0(r,\theta)\\
    &+\sum_{n=-\infty}^{\infty} \sum_{\substack{\alpha={K,K'} \\ \bar{\alpha}\ne\alpha} }c_\alpha \frac{i^n}{2} \Big[(S_{n,\alpha\alpha}-1)\psi_{n,\alpha}^{(\text{out})}(r,\theta)\ket{\alpha}\\
    & +S_{n,\alpha\bar{\alpha}} \psi_{n,\bar{\alpha}}^{(\text{out})}(r,\theta)\ket{\bar{\alpha}}\Big],
\end{split}
\label{Eq:Psi_Out}
\end{equation}
and the total wave function is obtained by,
\begin{equation}
     \Psi(r,\theta)=\sum_{n=-\infty}^{\infty}\left[ \Psi_n^<(r,\theta)+\Psi_n^>(r,\theta)\right].
\end{equation}

\section{Valley Birefringence}
\label{Sec:Valley}

Partially subjecting a graphene sheet to a gate potential that reverses its carriers character from electrons to holes between gated and non-gated regions leads to many interesting analogies between its electron dynamics and optical phenomena~\cite{Veselago-GrapheneTheory,cserti2007caustics,Asmar2013-Circular} The key ingredient to this phenomena is the reversal of the group velocity of quasiparticles between the regions with and without gate potentials. For example, if we consider a circularly gated region on graphene (Fig.~\ref{fig:system} with $V_0$ and taking $\Delta=0$), then, in the outer region, $r>R$, a quasiparticle's group velocity is $v^>_g=v_f(k_{x,r>R}\hat{x}+k_{y,r>R}\hat{y})/| k_{r>R}|$, while in the inner region we have a negative group velocity  $v^<_g=-v_f(k_{x,r<R}\hat{x}+k_{y,r<R}\hat{y})/| k_{r<R}|$ , here $ k_{r<R}$ ($ k_{r>R}$) is the wavevector in the inner (outer) region.  
The reversal of the group velocity from the outer to the inner region indicates that the gated region will act, in the semiclassical limit, as a circular electronic lens with a negative index of refraction $n=- k_{r<R}/k_{r>R}$, where $k_{r<R}$ is the wave number inside the gated patch and $k_{r>R}$ outside, and $n$ is deduced from the electronic Snell's law~\cite{graphene,Veselago-GrapheneTheory}. As shown in Fig.~\ref{fig:system}, in the limit $kR\gg 1$, the negative index of refraction leads to constructive interference between the different partial wave components and results in a probability density that forms cardioid caustics and cusps~\cite{cserti2007caustics}, in what mimics the optical caustics which arise from light refraction through a shaped medium and belong to a class of cusps in catastrophe theory~\cite{BERRY1980257}. Using differential geometry \cite{cserti2007caustics}, the positions of the cusps for each $p-1$ internal reflections can be shown to be
\begin{figure*}
    \centering
    \includegraphics[width=0.95\textwidth]{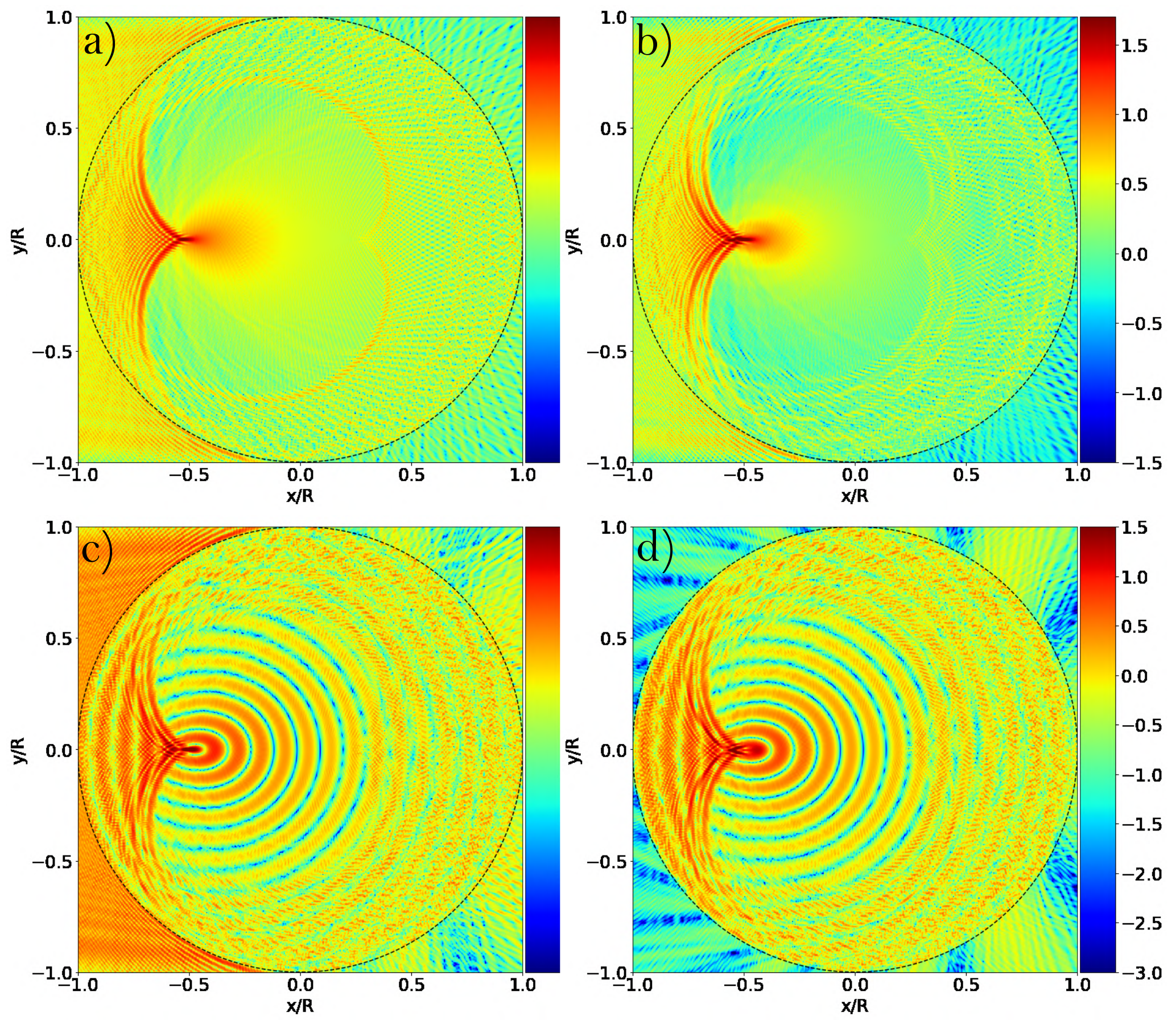}
\caption{Space dependence of the probability density (in $\log_{10}$ scale), for an incoming electron flux in the \text{x}-direction with valley polarization $K'$ and $kR=300$. The dashed line shows the boundary between the scattering regions. A gate potential $V_0R/(\hbar v_f)=600$ is present in the inner region for, a) pristine graphene, b) Kek-Y distorted graphene in the $r<R$ region with $\Delta=0.1$. Panel c) shows the valley-preserving component and d) the valley-flip component, for the case described in b).}
\label{fig:scatteringpicture}
\end{figure*}
\begin{equation}
    x_{cusp}(p)=\frac{(-1)^p}{|n|-1+2p}R,
    \label{Eq:Cusp}
\end{equation}
and in the case shown in Fig.~\ref{fig:scatteringpicture} a), as the amplitude decreases with each internal reflection, we can clearly distinguish the cusps corresponding to $p=1,2$.

If in addition to the gate potential the  circular region contains the Kekulé bond texture, then the electronic bands in this region will be characterized by $E_{\pm}=\pm\hbar v_f(1\pm \Delta)|\bm k| +V_0$. Therefore, the gating of this region leads to the  Fermi level intersecting the two degenerate hole bands, which are characterized by the two group velocities, $v_{g,\pm}=-v_f(1\pm \Delta)$. Then, when $\Delta\ne 0$, in addition to the sign reversal of the group velocity between both regions we also have the two different group velocities in the inner region. Hence, the Kekulé patterned and gated region will act as a circular lens with two negative indices of refraction
\begin{equation}\label{refindex}
    n_{\pm}=- \frac{k_{\pm,r<R}}{k_{r>R}},
\end{equation}
with $k_{\pm,r<R}=k_+,$ $k_-$ and are given in Eq.~\eqref{kpkm}. As shown in Fig.~\ref{fig:scatteringpicture}b) The Kekulé patterning of the circular region results into the doubling of the cusps and caustics of the circular lens, which reflects its birefringent nature. The degree of birefringence can be characterized by $\zeta=|n_+-n_-|$, and for the set of parameters in Fig.~\ref{fig:scatteringpicture}b) we get $\zeta\approx 0.25$. Moreover, the cusps locations is now modified to 
\begin{equation}
    x^{\pm}_{cusp}(p)=\frac{(-1)^p}{|n_\pm|-1+2p}R,
    \label{Eq:Cusp2}
\end{equation}
and the spatial separation between the two cusps is found by $|x^{+}_{cusp}-x^{-}_{cusp}|$. In Fig.~\ref{fig:scatteringpicture} c) we show the valley preserving amplitude component, $|\psi_{KK}(r)|^2$, and Fig.~\ref{fig:scatteringpicture} d) the valley mixing component, $|\psi_{KK'}(r)|^2$. From these figures we can notice that the Kekulé bond texture leads to the oscillation of the valley component as electrons travel in the patterned region, in what mimics the electron's spin-momentum coupling in the presence of a Rashba interaction~\cite{Rashba,Asmar2013-Circular}.

\section{Low Energy Scattering}
\label{Sec:Low}
The scattering process can be further analyzed by obtaining the different types of cross sections, such as the total cross section $\sigma_t$ which tells us the magnitude of the interaction between the incoming flux and the scattering region, the transport cross section $\sigma_{tr}$ that describes the average momentum transfer during the scattering, and the skew cross section $\sigma_{sk}$ which shows the asymmetry in the scattering around the axis of the incoming flux. This quantities can be obtained through the scattering amplitude $f(\theta)$, which can be found in the far field limit, {\it i.e.}, via the asymptotic form of the wave function as $r \rightarrow \infty$
\begin{equation}
    \Psi(r \rightarrow \infty)\rightarrow \Phi_0+\sum_{n}\sum_{\alpha,\beta}c_{\alpha}f_{n,\alpha\beta}(\theta)\frac{e^{ikr}}{\sqrt{r}}\ket{\beta},
    \label{Eq:Asymptotic}
\end{equation}
and using the asymptotic expansion of the Hankel functions,
\begin{equation}
    H_n(kr)^{(1)/(2)}\rightarrow \sqrt{\frac{2}{\pi k r}}e^{\pm i(kr-\frac{n \pi}{2}-\frac{\pi}{4})},
\end{equation}
by comparing  Eq.~\eqref{Eq:Psi_Out} to Eq.~\eqref{Eq:Asymptotic} we can deduce the scattering amplitude for each  partial wave component in terms of the $S$ matrix components 
\begin{equation}
f_n=\frac{e^{-i \pi/4}}{\sqrt{2 \pi k}}
    \begin{bmatrix}
            S_{n,K',K'}-1 & -i S_{n,K',K}\\
            i S_{n,K,K'} & S_{n,K,K}-1
    \end{bmatrix}
    ,
\end{equation}
where $S_{n,\alpha\beta}$ are the valley preserving $(\alpha=\beta)$ and valley mixing scattering $(\alpha\ne\beta)$ matrix elements corresponding to the $n^{\rm th}$ partial wave component ($\alpha$ and $\beta$ represent the Dirac points, either $K$ or $K'$). Then, for each process (valley preserving and valley mixing), we find the corresponding differential cross section,
\begin{equation}
    \sigma_{\alpha \beta}(\theta)= \left| \sum_{n=-\infty}^\infty f_{n,\alpha\beta} e^{in\theta}\right|^2,
\end{equation}
total cross section,
\begin{equation}
    \sigma_{t,\alpha \beta}=\int_{-\pi}^{\pi}\sigma_{\alpha\beta}(\theta) d\theta=2\pi \sum_{n=-\infty}^{\infty} |f_{n,\alpha\beta}|^2,
\end{equation}
transport cross section,
    \begin{equation}
    \begin{split}
        \sigma_{tr,\alpha\beta}&=\int_{-\pi}^{\pi}\sigma_{\alpha\beta}(\theta)(1-\text{cos}\theta) d\theta\\
        &=\sigma_{t,\alpha\beta}-2\pi\sum_{n=-\infty}^{\infty} \text{Re}(f_{n,\alpha\beta} f_{n+1,\alpha\beta}^*),
        \end{split}
\end{equation}
and the skew cross section,
    \begin{equation}
    \begin{split}
        \sigma_{sk,\alpha \beta}&=\int_{-\pi}^{\pi}\sigma_{\alpha\beta}(\theta)\text{sin}\theta d\theta\\
        &=2 \pi \sum_{n=-\infty}^{\infty} \text{Im}(f_{n,\alpha\beta} f_{n+1,\alpha\beta}^*),
        \end{split}
\end{equation} 
by summing over all different allowed processes
\begin{equation}
    \sigma_\eta=\sum_{\alpha,\beta} \sigma_{\eta,\alpha \beta},
\end{equation}
we obtain the total, transport, and skew cross sections ($\eta \in \{t,tr,sk\}$).

\begin{figure}[!htbp]
\begin{center}
{\includegraphics[width=.48\textwidth]{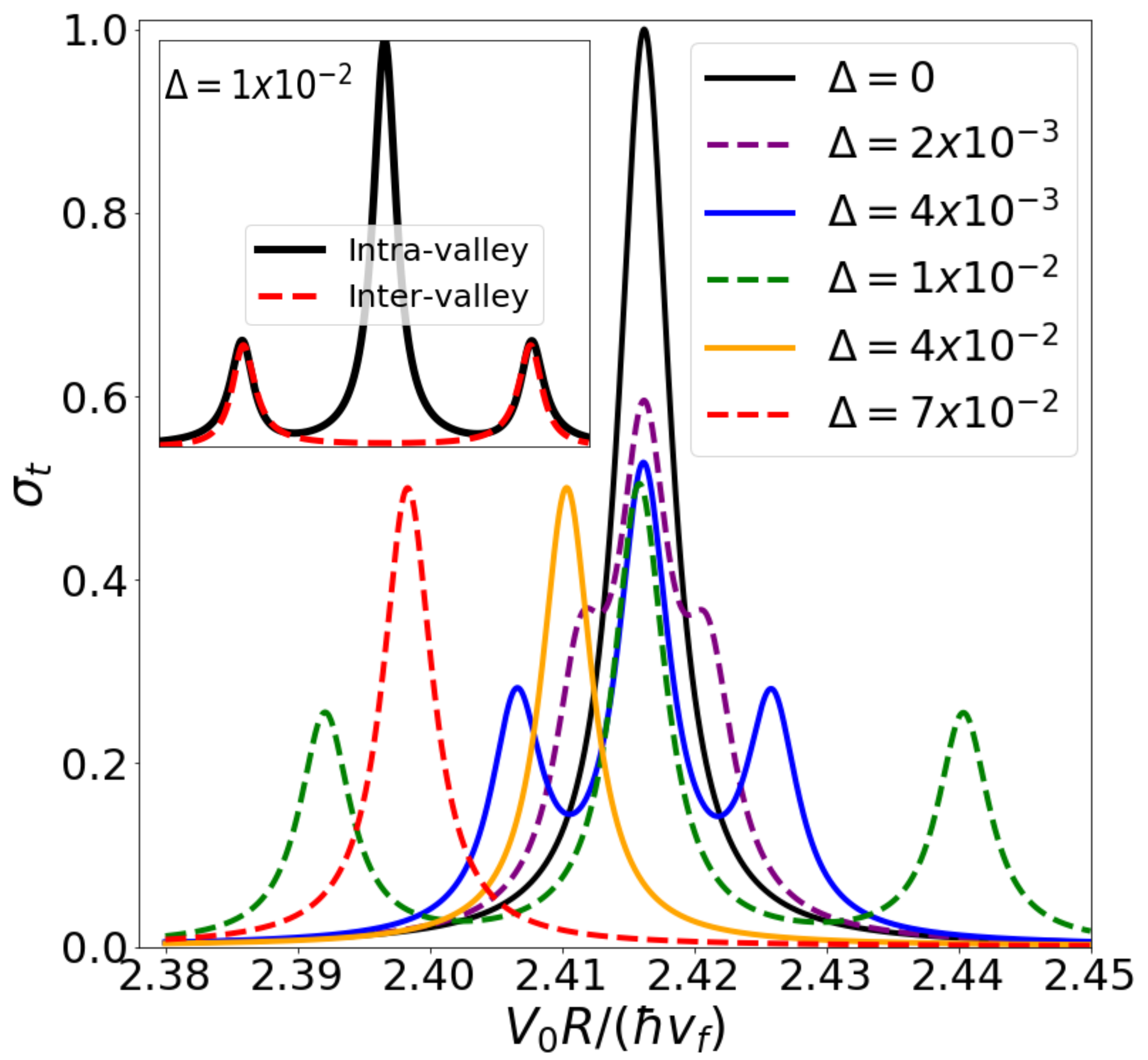}}
\caption{Total cross section $\sigma_t$ as a function of $V_0$ for incoming electrons in the $x$-direction with $kR=1.5\times10^{-3}$. 
(Inset) Total cross section for intra-valley  $\sigma_{t,KK}+\sigma_{t,K'K'}$ and inter-valley $\sigma_{t,KK'}+\sigma_{t,K'K}$ process with $\Delta=0.01$.}
\label{fig:TotCS}
\end{center}
\end{figure}

For low carrier concentrations and small regions with Kekulé bond texture ($kR \ll 1$) the most significant scattering channels are those of angular momentum $n=-1,0,1$. Within this regime, we show in Fig.~\ref{fig:TotCS} the total cross section against the strength of gate potential $V_0$. 
In the absence of Kekulé patterning, the total cross section of the gated region displays one peak which uniquely arises from the valley preserving process and indicates the formation of quasi-bound states in this region with finite life-time characterized by the width of the peak~\cite{Isotropic,asmar2015symmetry}. An increasing strength of the Kekulé interaction leads to the central (valley-preserving) peak height shrinking and its location shifting, while two new resonant (valley-mixing) peaks emerge. These two new peaks correspond to quasi-bound states forming due to valley mixing processes as it is shown in the inset of Fig.~\ref{fig:TotCS}, and consequently their height increases with increasing values of $\Delta$, as shown in Fig.~\ref{fig:TotCS}.

When local interactions in a graphene sheet lead to the breaking of effective time reversal (time reversal per valley) while preserving the total time reversal, as is the case for the Kekulé patterning, it is possible to have a skew scattering, and by symmetry considerations it can be shown that~\cite{asmar2015symmetry}
\begin{subequations}
\begin{equation}
     \sigma_{sk,\alpha \alpha}=- \sigma_{sk,\bar{\alpha} \bar{\alpha}},
\end{equation}
\begin{equation}
    \sigma_{sk,\alpha  \bar{\alpha}}=0.
\end{equation}
\end{subequations}  
The latter equations indicate that electrons with opposite valley polarization get deflected towards opposite directions as they get scattered, thus producing a valley-Hall effect.
To measure the asymmetry of the scattering per valley we calculate the skew parameter $\gamma_V$, which is defined as,
\begin{subequations}
\begin{equation}
    \gamma_V=\frac{1}{2}(\gamma_K-\gamma_{K'}),
\end{equation}
where
\begin{equation}
    \gamma_\beta=  \frac{\sum_\alpha \sigma_{sk,\alpha \beta}} {\sum_\alpha \sigma_{tr,\alpha \beta}},
\end{equation}
\end{subequations}
this quantity is directly connected to the transverse valley currents and is equal to the valley Hall angle at zero temperature in the absence of side-jump effects \cite{ferreira2014extrinsic},
\begin{equation}
    \Theta_{V H}=\frac{j_{V H}}{j_x}=\gamma_V.
\end{equation}
\begin{figure}[!htbp]
\begin{center}
{\includegraphics[width=.48\textwidth]{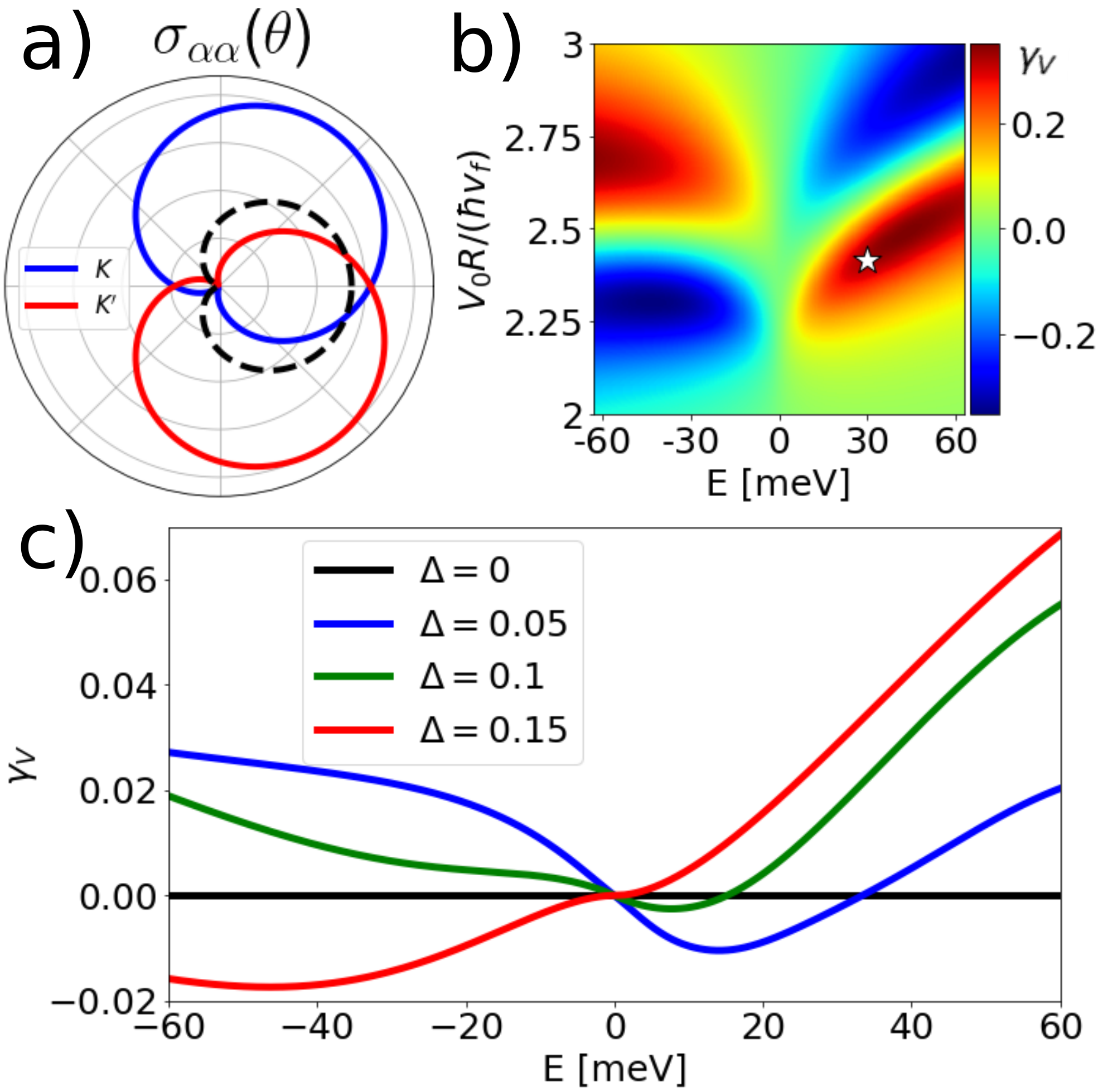}}
\caption{a) Differential cross section for valley preserving processes, $K$-valley (blue) and $K'$-valley (red), showing the tilt of electrons with opposite valley-polarization towards opposite directions around the $x$-axis. The dashed black line in a) corresponds to the differential cross section without Kekulé distortion. b) Valley skew parameter $\gamma_V$ as function of both energy and the gate potential for a region of $R=9\text{\AA}$ and Kekulé amplitude $\Delta=0.1$, the star marker indicates the values used for a). c) Average of $\gamma_V$ as a function of energy for $4000$ randomly sized Kek-Y regions $(9\leq R \leq18$) \AA, considering $V=1$ eV and different values of $\Delta$.}
\label{fig:Skew}
\end{center}
\end{figure}

In the presence of the Kekulé-Y modulation, the valley asymmetry of scattering around the $x$-axis can be also deduced from the valley dependent differential cross section. In Fig~\ref{fig:Skew} a), we  present the differential cross section per valley for the set of parameters indicated by a star marker in Fig.~\ref{fig:Skew} b). In contrast, we notice a symmetric scattering in the absence of the Kekulé-Y modulation, which is shown by the dashed black line in Fig~\ref{fig:Skew} a). To show the dependence of the skew scattering in our system on the local potential of the Kekulé-Y patterned patches $V_0$, and the Fermi energy ($E$), in Fig. \ref{fig:Skew} b) we show a map of the skew parameter, $\gamma_V$, as a function of $V_0$ and $E$ for Kekulé patterned regions with $R=9$ \AA $ $ and $\Delta = 0.1$. In the latter we should note that the regions of high $\gamma_V$ coincide with the regions of resonant scattering, {\it i.e.}, resonant regime in the total cross section (Fig.~\ref{fig:TotCS}), and which indicates that skew scattering is resonantly enhanced~\cite{ferreira2014extrinsic}. To demonstrate the robustness of skew scattering in the system to variations in size of the Kekulé-Y patterned patches, we consider a uniform random distribution of impurity sizes in the dilute limit. In Fig.~\ref{fig:Skew} c) we show the average of $\gamma_V$ for different values of $\Delta$ and $V_0$. Since skew scattering is resonantly enhanced, then its detection survives the random variations in the sizes of the Kekulé patterned patches in the  dilute limit, and which allows for the detection of valley Hall effect signatures in transport experiments. We also note that since the $RV_0/(\hbar v_f)$ governs the appearance of the different scattering regimes in Fig.~\ref{fig:TotCS}, then skew scattering is also robust to variations in the locally enhanced potential.

\section{Conclusions}
We have studied the scattering of Dirac Fermions from Kekul\'e distorted and gated regions in graphene. For large Kekul\'e patterned and gated regions, we have shown that the scattering of electrons from these circular patches leads to the formation of caustics and cusps reminiscent of a circular birefringent electronic lens with two negative indices of refraction. Moreover, the separation of the cusps in the circular lens is proportional to the Kekul\'e interaction and provides a direct measure of its strength in systems with tailored Kekul\'e patches.

For low carrier concentrations, we have shown that the presence of scatterers with a locally enhanced Kekul\'e interaction and gate potential leads to the electrons from different valleys deflecting in opposite directions due to the skew scattering that is enhance by the Kekul\'e distortion. Skew scattering in the system, leads to the appearance of a valley Hall effect. We have also shown that the skew scattering-generated valley Hall effect can be detected in systems where the Kekul\'e patterning is not perfect, but it instead leads to the formation Kekul\'e patches of random sizes and potentials. The latter suggests the plausible experimental realization and detection of the skew scattering induced valley Hall effect in Kekul\'e patterned graphing systems via four probe experiments.

\section{Acknowledgments}

E.A. and R.C.B. acknowledges useful discussions with Alex Santacruz. This work was supported by UNAM DGAPA PAPIIT IN102620 (E.A. and G.G.N.), CONACyT project 1564464 (E.A. and G.G.N.), and the National Science Foundation via Grant No. DMR-2213429 (M.M.A.).

\appendix
\section{Boundary Conditions}
In this appendix we find explicit solutions for the coefficients in Sec.~\ref{Sec:scattering}, which are found by solving for the boundary conditions. The solution of the system of equations resulting from the boundary condition $\Psi_n^<(R,\theta)=\Psi_n^>(R,\theta)$ gives us the following analytical expressions for the $S_n$ matrix elements and the amplitudes $T_n^\pm$,
\begin{subequations}
\begin{equation}
\begin{split}
    S_{n,K'K'}=&ss'(H_{n}^{(1)}H_{n-1}^{(2)}X_n+H_{n+1}^{(1)}H_{n}^{(2)}X_{n-1})/D_n\\
    &-(2H_{n+1}^{(1)}H_{n-1}^{(2)}Q_n+H_n^{(1)}H_n^{(2)}Z_n)/D_n,
\end{split}
\end{equation}
\begin{equation}
\begin{split}
    S_{n,KK}=&ss'(H_{n-1}^{(1)}H_{n}^{(2)}X_n+H_{n}^{(1)}H_{n+1}^{(2)}X_{n-1})/D_n\\
    &-(2H_{n-1}^{(1)}H_{n+1}^{(2)}Q_n+H_n^{(1)}H_n^{(2)}Z_n)/D_n,
\end{split}
\end{equation}
\begin{equation}
    S_{n,K'K}=\frac{-ss'Y_{n}P_{n}}{D_n},
\end{equation}
\begin{equation}
    S_{n,KK'}=\frac{-ss'Y_{n-1}P_{n+1}}{D_n},
\end{equation}
\begin{equation}
\begin{split}
    T_n^+=&c_1(j_{n+1}^-H_n^{(1)}-ss'j_n^-H_{n+1}^{(1)})P_n/D_n\\
    &+c_2(j_{n-1}^-H_n^{(1)}-ss'j_n^-H_{n-1}^{(1)})P_{n+1}/D_n,
\end{split}
\end{equation}
\begin{equation}
\begin{split}
    T_n^-=&c_1(j_{n+1}^+H_n^{(1)}-ss'j_n^+H_{n+1}^{(1)})P_n/D_n\\
    &-c_2(j_{n-1}^+H_n^{(1)}-ss'j_n^+H_{n-1}^{(1)})P_{n+1}/D_n,
\end{split}
\end{equation}
\end{subequations}
where we defined,
\begin{subequations}
\begin{equation}
\begin{split}
    D_n=&-ss'(H_{n}^{(1)}H_{n-1}^{(1)}X_n+H_{n+1}^{(1)}H_{n}^{(1)}X_{n-1})\\
    &+2H_{n+1}^{(1)}H_{n-1}^{(1)}Q_n+H_n^{(1)}H_n^{(1)}Z_n,
\end{split}
\end{equation}
\begin{equation}
    X_n=j_{n}^+j_{n+1}^-+j_{n+1}^+j_{n}^-,
\end{equation}
\begin{equation}
    Y_n=j_n^+j_{n+1}^--j_{n+1}^+j_n^-,
\end{equation}
\begin{equation}
    Z_n=j_{n-1}^+j_{n+1}^-+j_{n+1}^+j_{n-1}^-,
\end{equation}
\begin{equation}
    Q_n=j_{n}^+j_{n}^-,
\end{equation}
\begin{equation}
    P_n=H_{n}^{(1)}H_{n-1}^{(2)}-H_{n-1}^{(1)}H_{n}^{(2)},
\end{equation}
\end{subequations}
here all Hankel functions are evaluated at $kR$ and $j_n^\pm=J_n(k_\pm R)$.
\bibliographystyle{unsrt}
\bibliography{refs.bib}
\end{document}